\documentclass[onecolumn,aps,pre]{revtex4-1}
\usepackage{amsmath,amsfonts}
\usepackage[utf8]{inputenc}
\usepackage{setspace}
\usepackage{caption}
\usepackage{color}
\usepackage{subcaption}
\usepackage{graphicx}
\usepackage{color}

\begin{document}
\title{Simulation of Spatial Systems with Demographic Noise}
\author{Haim Weissmann, Nadav M. Shnerb, David A. Kessler}
\affiliation{Dept. of Physics, Bar-Ilan University, Ramat-Gan 52900 Israel}
\begin{abstract}
Demographic (shot) noise in population dynamics scales with the square root of the population size. This process is very important, as it yields an absorbing state at zero field, but simulating it, especially on spatial domains, is a non-trivial task. Here we compare the results of two operator-splitting techniques suggested for simulating the corresponding Langevin equation, one by  Pechenik and Levine (PL) and the other by Dornic, Chat\'e and Mu\~noz (DCM). We identify an anomalously strong bias toward the active phase in the numerical scheme of DCM, a bias  which is not present in the alternative scheme of PL. This bias strongly distorts the phase diagram determined via the DCM procedure for the range of time-steps used in such simulations.  We pinpoint the underlying cause in the inclusion of the diffusion, treated as an on-site decay with a constant external source, in the stochastic part of the algorithm. Treating the diffusion deterministically is shown to remove this unwanted bias while keeping the simulation algorithm stable, thus a hybrid numerical technique, in which the DCM approach to diffusion is applied but the diffusion is simulated deterministically, appears to be optimal.
\end{abstract}
\maketitle

\section{Introduction}

Systems with demographic noise play an important role in many different problems, including birth-death processes, ecology, population genetics, reaction-diffusion processes, infection models, etc.~\cite{Durrett1994}. Demographic noise is a very important factor in the dynamics of these systems, since the noise vanishes at zero abundance, rendering the inactive state an absorbing state.  Technically, demographic stochasticity is a special kind of multiplicative noise, proportional to the square-root of the fluctuating field.  Thus, at small amplitudes, the noise is typically dominant. As a result, simulating the dynamics of a system with demographic stochasticity requires special care; it is crucial, for instance, that the noise not cause the amplitude to become negative~\cite{Dickman1994}, which would occur with a naive treatment of the noise.

These problems are especially severe in spatial systems, where many local populations are connected by diffusion of individuals. There have been numerous schemes proposed to make the simulation of a spatially extended system  more efficient. 
One approach is to simulate an individual-based model using  a Gillespie-type algorithm~\cite{Gillespie1976}, where the rate for each event is taken into account and the next event is picked from the complete set of probabilities. If the demographic noise, arose, as is typical, from the original model having discrete individuals, then this is an exact treatment of the problem. However,  for large spatial systems this procedure is prohibitively expensive, especially at large population levels.  Having to pick the next event from the number of different processes that can occur, each with its own unique rate, slows down the calculation enormously.  Alternatively, one may still use an individual-based model but discretize the process in time, along the lines of how one simulates a deterministic partial differential equation.  In this scheme~\cite{Kessler1998} the events on different sites occur in parallel - the number of births, for example, at a given site, is given by a Poisson or Binomial deviate, with the mean determined by the birth rate times the time step -  and so this is much more efficient than Gillespie, especially at large density (small noise).  This is still not ideally efficient, in that a different random deviate must be generated at every site for every different process. Furthermore, being essentially an Euler type scheme, there are strong restrictions on how large the time step can be.  The finite step involved in this procedure perforce limits the accuracy, but this source of error is typically overwhelmed by the stochastic sampling error.

Besides the individual-based approach, one can instead use a continuum approach, either Fokker-Planck or Langevin~\cite{vanKampen1992}.  If the model under study is such a continuum equation, such as the stochastic Fisher equation, then the individual-based models discussed above are only an approximation.  Otherwise, the continuum equation is a certain large-population limit of the original model. In this continuum approach,  the strength of the demographic noise is given by a single parameter, the amplitude of the square root noise. Here, one has to face squarely the difficulties inherent is the square-root type multiplicative noise.  One noteworthy approach was that of  Pechenik and Levine (PL)~\cite{Pechenik1999}, who introduced an operator splitting method, based on an analytic solution of the pure birth-death process.  In a single time-step of the PL simulations, the birth-death process is simulated exactly, followed by a deterministic update implementing the diffusion and nonlinear reaction terms   This method has the possibility of being stable for all time step sizes. Using this method, PL were able to simulate the Fisher process at very small noise levels, thereby confirming the analytical predictions of Brunet and Derrida~\cite{Brunet1997} for this limit.  

In a subsequent development, Dornic, Chat\'e and Mu\~oz~\cite{Dornic2005} (DCM) proposed a more efficient variation on this split-step scheme and used it to study a variety of systems. The aim of this paper is to point out a (quite delicate) problem in the procedure suggested by DCM, and to explain and demonstrate some of its implications.    

The method of DCM differs from the PL algorithm in two aspects. 
 \begin{enumerate}
   \item  The analytical solution that PL used for the local birth-death process  is given by a Bessel function and it is inconvenient to generate deviates for this distribution.  Dornic, et al. succeeded in showing that this distribution can be expressed as a convolution of a Poisson process with a Gamma distribution deviate.  Since efficient implementations of both these random number generators exist, the simulation of the process has thereby been rendered much simpler.
   \item In addition, DCM have modified the treatment of the diffusion term, incorporating it approximately in the stochastic process.  They split the diffusion operator into two pieces: an  on-site sink term, which causes the on-site density to decay at a rate proportional to the diffusion constant $D$, and an  off-site source term, accounting for the incoming flux of particles from the neighboring sites.  Given that the source is taken as fixed for the duration of the time step, each site still evolves independently. 
 \end{enumerate}
 
The DCM scheme also involves an  operator-splitting with time-step $\Delta t$, but with a deterministic update only for the nonlinear terms, following the local noise-diffusion processes which generates new stochastic values of the concentration field at each site.  The new values of the concentration then determine the diffusive source terms for the next time step of $\Delta t$. Formally, both the PL as well as the DCM method are first order methods, with errors proportional to  $(\Delta t)^2$.  The advantage of treating the diffusion using the DCM technique, instead of simulating it numerically as PL did, is that the DCM method  is guaranteed to be stable no matter the size of $\Delta t$.  This is crucial if the system exhibits a long equilibration time, as it will do in the vicinity of a phase transition.  This stability can be achieved also within the PL scheme, if the deterministic diffusion step is implemented in a stable way, say by means of a Crank-Nicholson approach~\cite{NumRec}, at the cost, however, of added complexity and running time.
  
Thus it appears that the two methods are basically equally efficient, with an edge to the DCM  scheme.  The first indication that there is something untoward occurring in the DCM procedure appeared in a recent work of Martin, et al.~\cite{Martin2015}, who used this method to study catastrophic shifts in a two dimensional model of a population with positive feedback (strong Allee effect). Deterministically this system supports two stable fixed points, one at zero field and the other at some finite value. On spatial domains, one of these states invades the other, unless the external parameters (like, e.g., the stress affecting the population) are tuned to the stall (Maxwell) point.  The  numerical results of Martin et al.~\cite{Martin2015}  (see Fig. 2) indicated that in the stochastic system the active state invades the absorbing state even on the high-stress side of the Maxwell point, i.e., in the regime where in the deterministic system the  inactive state is dominant.  If true, this would be  a very surprising feature, since the demographic noise destabilizes the active state, but can't touch the inactive state where its amplitude (which is proportional to the square root of the concentration) vanishes.  Moreover, DCM presented an analytic calculation showing that the effect of the noise is to push the system away from the first-order transition toward a continuous transition, as always occurs in one dimension, and  the continuous directed-percolation transition point must be at lower stress than the Maxwell point~\cite{Weissmann2014}.

In this paper, then, we investigate this anomalous aspect of the DCM scheme, showing that while indeed the DCM converges to the correct answer in the $\Delta t \to 0$ limit, for finite $\Delta t$ the scheme gives rise to surprisingly large corrections that systematically favor the active state.  This turns out to be due to the subtle interplay of their incorporation of the off-site diffusional influx into the stochastic process within the split-step nature of their scheme.  Thus, to achieve qualitatively correct results on, e.g., the dependence of the stall transition on the noise, with the DCM scheme requires quite small time-steps. The PL scheme, on the other hand, which treats the diffusion deterministically, does not suffer from this problem, and so allows much larger time-steps for the same accuracy. When one is interested only in universal properties such as critical exponents, this issue with the DCM scheme does not cause problems.  However, much greater care is need when studying the locations of the transitions, which are not universal, and as above, can lead to qualitatively incorrect behavior. 

 We start with a short review of the two technique, then we will quantitatively compare the results of the two different scenarios first in the one-dimensional context where the physics is simpler than in two dimensions, and then in a two-site model.  This will afford us the needed insight into the origin of the DCM bias,
 and point to an optimally efficient and accurate algorithm.  We conclude by returning to the original ``scene of the crime" and show that while reducing $\Delta t$ drastically removes the anomaly in the location of the stochastic stall transition that provoked this study, the PL method yields much more quantitatively (and of course qualitatively) correct results with DCM's original choice of $\Delta t$.

\section{The Two Methods: PL and DCM}

In this section, we review the details of the two methods, in order to fix notation and nomenclature.  
The PL scheme consists of breaking up the Langevin equation into two pieces, a stochastic balanced birth-death process yielding the demographic noise and a deterministic piece containing the rest of the dynamics.  An elementary  time step  consists of a stochastic update of duration $\Delta t$, followed by a deterministic update of the same duration.  The stochastic update constitutes an exact solution of the local equation
\begin{equation}
\frac{\partial\phi}{\partial t}(x,t) = \sigma \sqrt{\phi(x,t)} \eta;  \qquad \langle \eta(x,t) \rangle = 0; \qquad \langle \eta(x,t) \eta(x',t') \rangle = \delta(x-x')\delta(t-t')
\label{eqBD}
\end{equation}
This equation is purely local in $x$ so $\phi$ at every discrete site can be updated independently. Following the stochastic and purely local updating of $\phi(x)$, the rest of the dynamics (diffusion, linear and nonlinear interactions) is simulated deterministically for $\phi(x)$ for duration $\Delta t$ and then the stochastic step is taken again. 
 
 As opposed to this  algorithm employed by PL, DCM realized that the stochastic update for a given site $i$ can be implemented via
first generating a Poisson deviate $Q_i$ with mean $\lambda\phi_i$ where
\begin{equation}
\lambda = \frac{2}{\sigma^2 \Delta t}
\end{equation}
and then generating a Gamma deviate $R$ with shape parameter $Q_i$ and scale unity. The new $\phi_i$ is then $R_i/\lambda_i$.   This is the first aspect of the DCM work and being an exact reformulation, does not cause problems. In fact, as there is no reason not to incorporate it into the PL method, in the following we do so without comment.
  
The deterministic step of the PL algorithm may be implemented for a time $\Delta t$ by whatever means is convenient, which can be Euler if $\Delta t$ is small enough, or alternatively Crank-Nicholson (or other stable techniques) if one wishes to increase $\Delta t$.  Since the split-step induces errors of order $\Delta t$, stability is more critical than accuracy for this part of the calculation.

The DCM scheme treats diffusion differently: this is the second aspect of the DCM method mentioned above, the one we considered to be  problematic.  DCM decompose the second-difference diffusion operator $D(\phi_{i+1}-2\phi_i+\phi_{i-1})/2(\Delta x)^2$, into two pieces, one a local decay term $-2D\phi_i/(\Delta x)^2$ and the second a source term, which is taken to be constant during the interval $(t_0,t_0+\Delta t)$. This gives rise to a different Langevin equation for the "linear step" (that may include linear growth/decay with rate $\alpha$, stochasticity and diffusion):
\begin{equation}
\frac{\partial\phi}{\partial t}(x,t) = (\alpha -\frac{2D}{(\Delta x)^2} )\phi(x,t) + S(x) + \sigma \sqrt{\phi(x,t)} \eta; \qquad  S(x)=\frac{D}{(\Delta x)^2}(\phi(x+\Delta x,t_0) + \phi(x-\Delta x,t_0))
\end{equation}
This (approximate) Langevin equation is also exactly solvable, and again involves generating a Poisson variate $Q_i$ at each site with mean $\lambda\phi_i\exp(\nu \Delta t)$ ,where now
\begin{equation}
\lambda = \frac{2\nu}{\sigma^2 (\exp(\nu\Delta t)-1)};   \qquad \nu=\alpha -2D/(\Delta x)^2,
\label{eqlambda}
\end{equation}
followed by a Gamma deviate, $R_i$ with shape parameter $2S_i/\sigma^2$.  The new $\phi_i$ is then again $R_i/\lambda$.  Any additional nonlinear dynamics is then implemented deterministically for an interval $\Delta t$, a la PL.

\section{A One-Dimensional System}

To see the anomalous behavior of the DCM scheme more clearly, we first study the model considered by Martin et al for  a one-dimensional  system, where the physics is simpler, since there the transition is always continuous. We compare the results of the DCM algorithm to that of PL.  The model  we choose to investigate is precisely the one-dimensional form of the model studied by Martin, et al. and is described by the Langevin equation
\begin{equation}
\frac{\partial\phi}{\partial t} = D\frac{\partial^2 \phi}{\partial x^2} + \alpha \phi + \beta \phi^2 - \gamma \phi^3 + \sigma \sqrt{\phi} \eta
\label{eqGL}
\end{equation}
where, as usual, $\eta$ is a unit-strength zero-mean white noise. 

Let us consider first the features of the deterministic dynamics, $\sigma =0$. $\alpha$ is the stress parameter in this model. If $\alpha$ is positive, the state $\phi=0$ is unstable and there is only one stable state at $\bar{\phi} = (\beta + \sqrt{\beta^2 + 4 \alpha \gamma})/(2\gamma)$. When $\alpha <0$ the zero state is stable, but the active state invades as long as $\alpha$ is above the Maxwell (stall) point, $\alpha_\textit{MP}  = -2\beta^2/9\gamma$. If $\alpha < \alpha_\textit{MP}$ the inactive phase invades the active one. Finally, below $\alpha_T = -\beta^2/4\gamma$ the active phase loses its stability and the only stable solution is at $\phi=0$. $\alpha_T$ is thus the tipping point, below which the deterministic active state collapses even without invasion of the inactive state (i.e., even the local deterministic dynamics does not support an active state).        

\begin{figure} 
\includegraphics[width=0.8\textwidth]{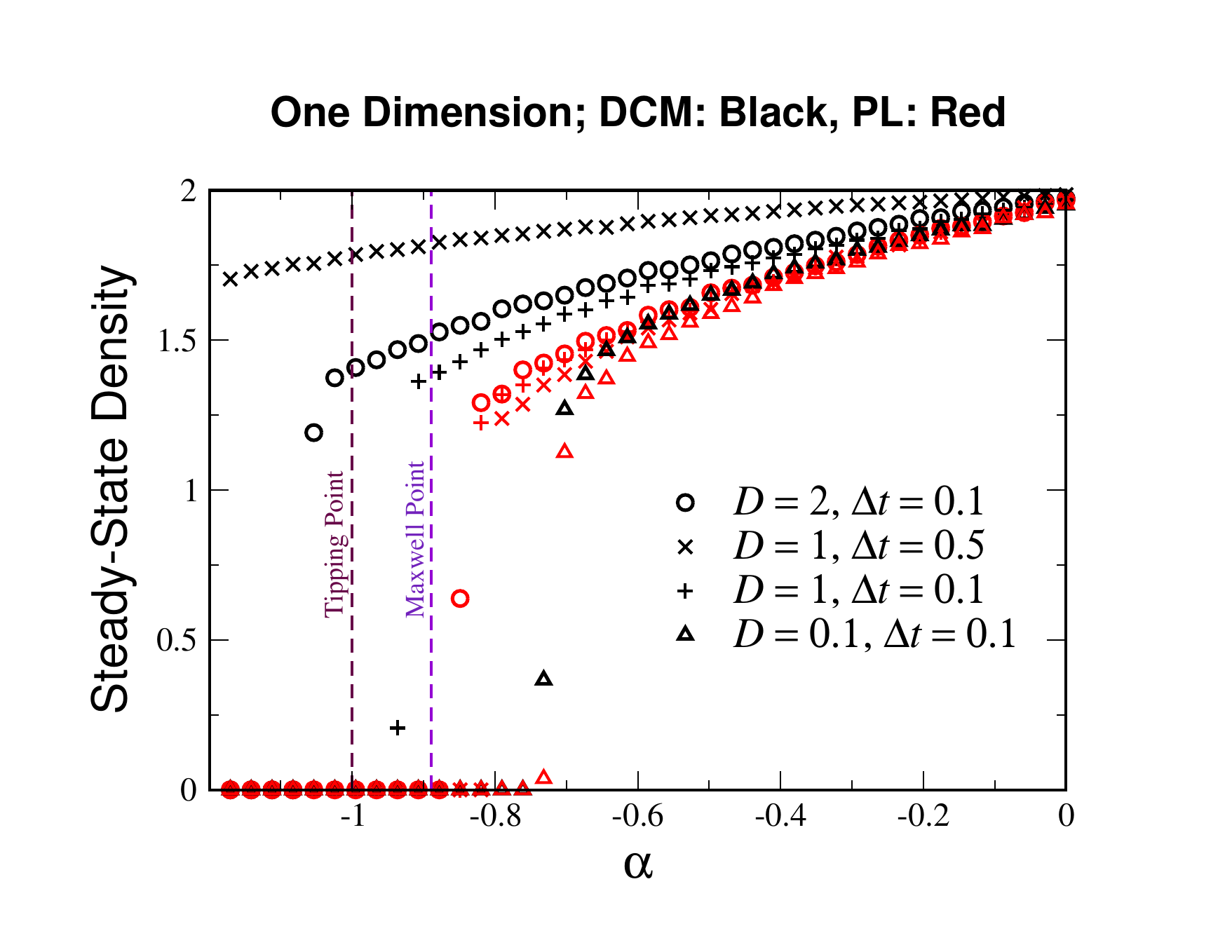}
\caption{Steady-state density as a function of the stress parameter $\alpha$ for the one-dimensional Ginzburg-Landau system of Eq. (\ref{eqGL}), with $\beta=2$, $\gamma=1$, $\sigma^2=0.2$, and various combinations of $D$ and $\Delta t$ as shown.  Data is shown in black for the DCM method and red for the PL  method (with diffusion handled via Crank-Nicholson). The values of $\alpha$ corresponding to the Maxwell point and the tipping point are indicated by dashed vertical lines.  The system size was $2^{10}$, with lattice space unity.}
\label{fig1D}
\end{figure}

In Fig. \ref{fig1D} we show the results of PL and DCM simulations, plotting the mean concentration
$\overline{\phi}$ as a function of $\alpha$ for fixed $\beta=2$, $\gamma=1$, $\sigma^2=0.2$ and various values of $D$ and $\Delta t$.   The most striking feature of this graph is that while, as expected, the transition according to the PL scheme is on the high $\alpha$ side of the Maxwell point, for larger values of $D$ and $\Delta t$, the location of the transition according to DCM is on the {\emph{low}} $\alpha$ side.  In fact, for $D=1$, $\Delta t = 0.5$, as well as $D=2$, $\Delta t = 0.1$, the transition according to the DCM scheme is on the low $\alpha$ side of the deterministic tipping point!  We see also from this graph that for given $D$, lowering $\Delta t$, lowers the mean concentration for both schemes, but the change in concentration is much larger for the DCM scheme. We will confirm later that both methods converge to the same answer as $\Delta t\to 0$.  The question is how misleading are the finite $\Delta t$ results.

\section{Two-Site Logistic System}

Which aspect of the DCM method is causing such a huge numerical bias toward the active state?    To simplify matters somewhat, we move to a logistic system at criticality by choosing $\beta$ negative and dropping the $\alpha$ and $\gamma$ terms in Eq. (\ref{eqGL}).  Moreover, we restrict ourselves to two sites only, with the diffusion operator $\pm D(\phi_2-\phi_1)$.  The large bias of the DCM procedure persists in this very simple system, as we shall see.

It should be noted that although in principle there is no true extinction in the DCM system, as the particle density is always strictly positive, in practice even the DCM procedure generates extinction.  When the populations on all sites is small, the divergence at the origin of the Gamma distribution for small index means that the densities rapidly become extremely small, and in fact underflow, generating true zeros in the simulation.

\begin{figure}
\centering
\includegraphics[width=0.8\textwidth]{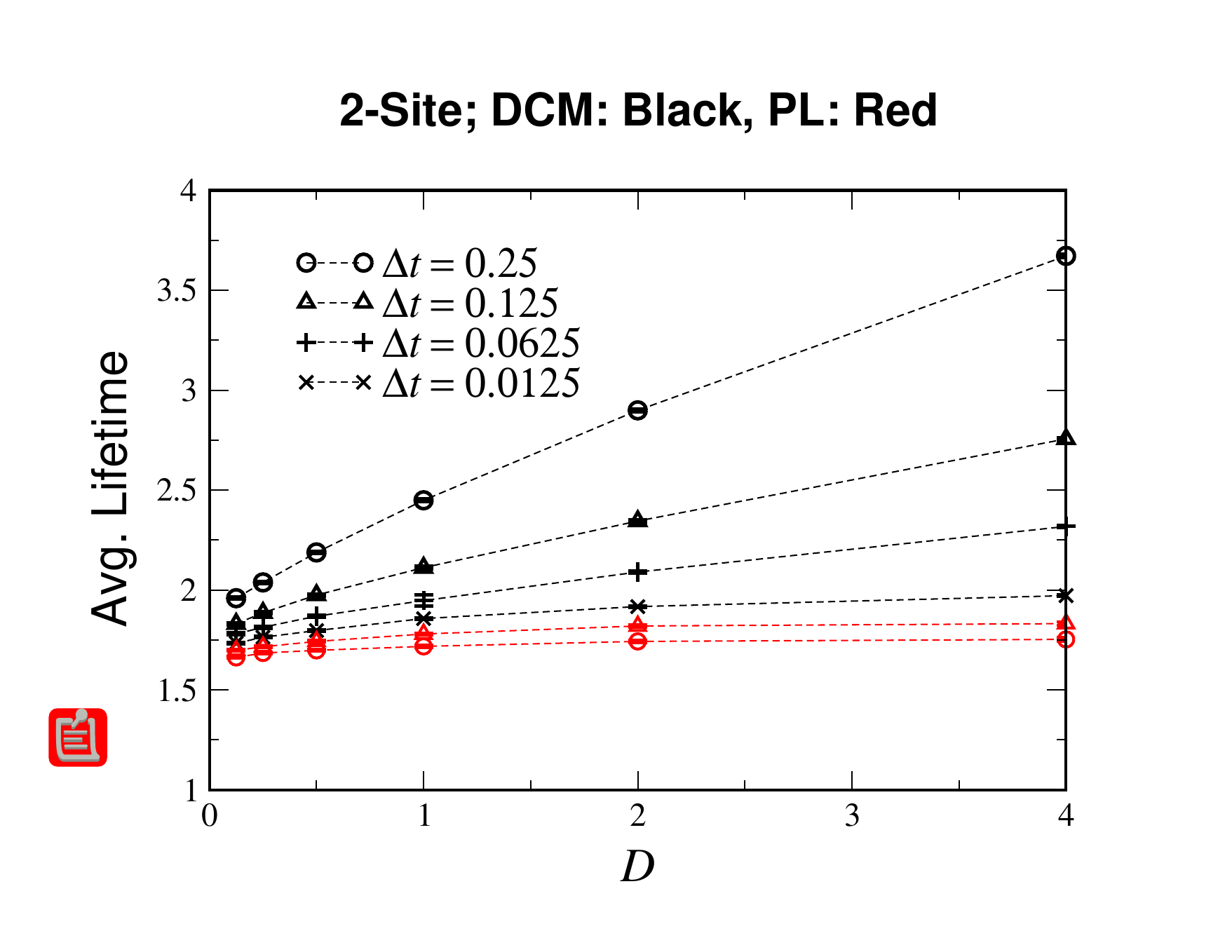}
\caption{The mean time to extinction (averaged over 50000 runs) of the two-site model with $\sigma^2=2$, $\beta=-1$, $\alpha=\gamma=0$ as a function of the diffusion constant  $D$ for various values of $\Delta t$.  Data is shown in black for the DCM method and red for the PL  method.  Error bars are $\pm 1$ std. error.}
\label{fig2sitelife}
\end{figure}

Still, there is a problem.  The data for the mean extinction time as a function of $D$ for various $\Delta t$ is shown in Fig. \ref{fig2sitelife} for the two methods.  We see again that PL has a slight bias favoring the inactive state, whereas DCM for $D=4$ has a roughly 20-fold larger bias toward the active state. The differences  between the two methods shrink with $D$ as expected, since the whole difference in the methods is in their treatment of diffusion.  Nevertheless, both methods are seen to converge to the same results as $\Delta t \to 0$.

To get a better handle on this phenomenon, we study the two-site model starting from its initial formulation as the continuum limit of a discrete particle birth-death process.   These are then the following basic processes:
\begin{align}
A_{1,2} &\stackrel{\alpha_0}{\to} 2A_{1,2}\nonumber\\
A_{1,2} &\stackrel{\alpha_0}{\to} 0\nonumber\\
2A_{1,2} &\stackrel{\beta_0}{\to} 0\nonumber\\
A_{1,2} &\stackrel{D}{\to} A_{2,1}
\end{align}
The first two lines represent a balanced birth-death process on the two sites, which gives rise to pure demographic noise.  The third line generates the logistic term, and the last is the diffusion. We can easily generate the master equation for this process:
\begin{align}
\frac{\partial P(n_1,n_2)}{\partial t} &= \alpha_0 (-(n_1+n_2)P(n_1,n_2) + (n_1+1)P(n_1+1,n_2) \nonumber\\
&{}\qquad\qquad + (n_1-1)P(n_1 - 1,n_2) + (n_2+1)P(n_1,n_2+1) + (n_2-1)P(n_1,n_2 - 1)) \nonumber\\
&{}+ \frac{\beta_0}{2}(-(n_1(n_1-1)+n_2(n_2-1))P(n_1,n_2) + (n_1+1)n_1P(n_1+1,n_2) + (n_2+1)n_2P(n_1,n_2+1)) \nonumber \\
&{}+ D(-(n_1+n_2)P(n_1+n_2) + (n_1+1)P(n_1+1,n_2-1) + (n_2+1)P(n_1-1,n_2+1)).
\label{master}
\end{align}
The continuum limit is obtained  by introducing a parameter $K \gg 1$, defining $x_{1,2}\equiv n_{1,2}/K$ and expanding $P$ to the second order to find  the desired Fokker-Planck equation,
\begin{align}
\frac{\partial P(x_1,x_2)}{\partial t} &= \alpha\left( \frac{\partial^2}{\partial x_1^2}(x_1P) + \frac{\partial^2}{\partial x_2^2}(x_2P)\right) + \beta\left(\frac{\partial}{\partial x_1}(x_1P) + \frac{\partial}{\partial x_2}(x_2P)\right) \nonumber \\
&{}+ D\left(\frac{\partial}{\partial x_1}-\frac{\partial}{\partial x_x}\right)((x_1-x_2)P),
\end{align}
with the ``bare" reaction rates related to the macroscopic ones by
\begin{equation}
\alpha_0 = \frac{K\sigma^2}{2}; \qquad \beta_0 = \frac{2\beta}{K}.
\end{equation}
Note that the diffusion constant does not get rescaled in this procedure.  It is also important to note that the logistic term and the diffusion term are deterministic in the continuum limit, since the second-order derivative terms in the expansion of these terms vanish in the large $K$ limit with our adopted scalings.

The essence of the DCM procedure is to break up the diffusion process into two pieces, one an on-site decay term and the second a source term from the neighboring site.  If one were to implement this, we would get the following master equation
\begin{align}
\frac{\partial P(n_1,n_2)}{\partial t} &= \alpha_0 (-(n_1+n_2)P(n_1,n_2) + (n_1+1)P(n_1+1,n_2) \nonumber\\
&{}\qquad\qquad + (n_1-1)P(n_1 - 1,n_2) + (n_2+1)P(n_1,n_2+1) + (n_2-1)P(n_1,n_2 - 1)) \nonumber\\
&{}+ \beta_0/2(-(n_1(n_1-1)+n_2(n_2-1))P(n_1,n_2) + (n_1+1)n_1P(n_1+1,n_2) + (n_2+1)n_2P(n_1,n_2+1)) \nonumber \\
&{}+ D(-2(n_1+n_2)P(n_1+n_2) + (n_1+1)P(n_1+1,n_2) + (n_2+1)P(n_1,n_2+1) + n_2P(n_1-1,n_2) + n_1P(n_1,n_2-1))
\end{align}
Going through the above procedure to generate the continuum Fokker-Planck equation, one obtains  {\emph{exactly}} the same result as before.  The only difference in principle is in the diffusive noise term, but as anyway the diffusive noise vanishes in the continuum limit, this  difference makes no difference.  Thus, while the DCM breakup of the diffusion term only conserves total particle number on average, as opposed to the true diffusive term, the added noise is in fact irrelevant in the continuum limit.

If the diffusional noise is not the issue, then what is?  There are two remaining differences between the schemes.  One is that PL treats the on-site and off-site diffusion terms on an equal basis, whereas DCM treats the on-site term as varying in the course of the time-step while the off-site term is kept fixed.  The second is that PL treats diffusion deterministically, whereas DCM incorporates diffusion into the stochastic process. Which then leads to the large errors in the DCM approach?  To answer this, we repeat our simulations for an intermediate model, where we handle the diffusion deterministically, as in PL, but using the
DCM breakup.  This leads to the deterministic diffusional update
\begin{equation}
\phi_1''  = \phi_1' e^{-D\Delta t} + \phi_2' \left(1 - e^{-D\Delta t}\right)
\end{equation}
where here $\phi_1'$ is the value of $\phi_1$ after the stochastic update, and $\phi_1''$ is the value after the diffusional update, with a similar equation for $\phi_2''$, as opposed to the exact update (for the diffusion alone)
\begin{equation}
\phi_1''  = \frac{1}{2}(\phi_1' + \phi_2')  +  \frac{1}{2}(\phi_1'- \phi_2' )e^{-2D\Delta t}.
\end{equation}
It is clear that the DCM-type scheme is accurate to first order in $\Delta t$ and conserves total particle number.  Upon redoing the numerical computation of Fig. \ref{fig2sitelife}, we obtain values which are within $1\%$ of those of the PL method, as compared to the up to $120\%$ difference between DCM and PL.  Thus it is clear that the problem in DCM does not come from the onsite-offsite representation of the diffusion operator, but rather it is the inclusion of  the diffusion in the stochastic process (at least as performed in the DCM method) that must be inducing of the large systematic bias toward the active state in DCM.

\begin{figure}
\centering
\includegraphics[width=0.8\textwidth]{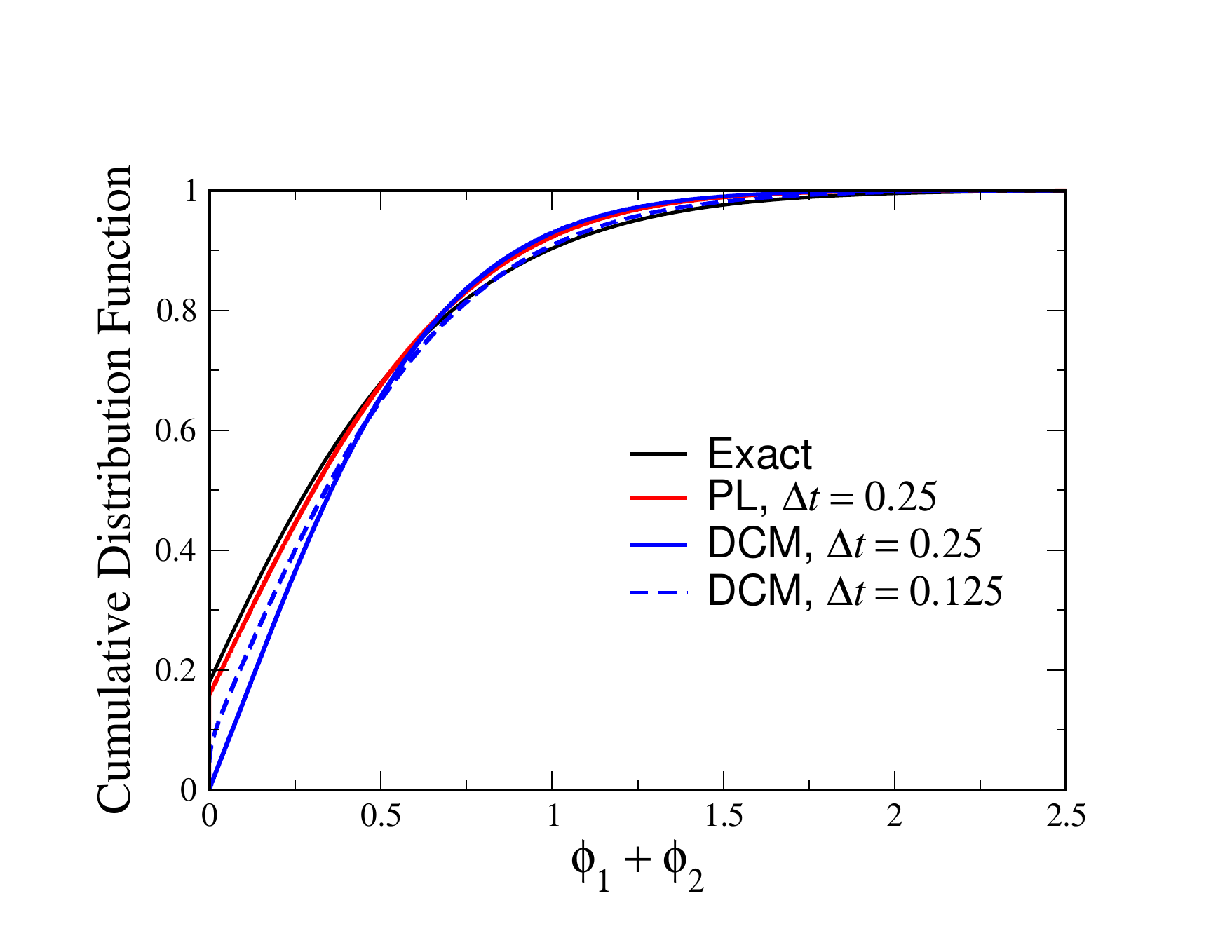}
\caption{The cumulative distribution function of the total particle number of the two-site model with $\sigma^2=2$, $\beta=-1$, $\alpha=\gamma=0$, $D=2$.  The exact results come from a numerical integration of the master equation, Eq. (\ref{master}) for the discrete process with $K=80$ and a truncation at 160 particles at each site.}
\label{figcumntot}
\end{figure}

 As the distinction between the methods manifests itself in the extinction statistics, we examine the extinction dynamics more closely.  Running either the PL and DCM simulations with a small $\Delta t$, and looking at those realizations that go extinct, we see that at time $t=0.25$ before extinction, the typical state has roughly equal numbers of particles on both sites, with the modal value being about $\phi_{1,2}=0.23$.  We then take a single step of $\Delta_t=0.25$ using both methods, as well as solving the full master equation to compare the distribution of the total number of particles in the system.  The results, shown in Fig. \ref{figcumntot} are striking. For values of $\phi_1+\phi_2$ not near zero, the cumulative probability distribution functions for the two methods are quite similar. The extinction story is however very different. For the exact dynamics, the probability of the system going extinct within $\Delta t$ is $18\%$.  Using a single step of size $\Delta t$ with PL gives around $16\%$ extinction, while DCM gives 0.  Taking two $\Delta_t=0.125$ DCM steps raises the true extinction rate to $0.03\%$, and an effective extinction rate (setting the extinction criterion to $\phi_1+\phi_2< 0.001$) to around $5\%$.   
 
 This phenomenon is due to a sort of ``rescue effect" present in the DCM.  The constant source in the DCM method means that the single step chance of one site being emptied is strictly zero if the neighboring $\phi > \sigma^2/2D$, since in this case, the shape parameter of the Gamma distribution is greater than unity (see the discussion after Eq. (\ref{eqlambda}) and so the  corresponding PDF vanishes at the origin.  In our case, $\phi=0.23$ is greater than this, but still the probability of the population at the site being less than $x$ scales as $x^{2D\phi/\sigma^2}$, so the probability that both sites are less than $x$ scales as $x^{4D\phi/\sigma^2}$, which in our case is essentially linear, yielding no extinctions (independent of the precise criterion). It takes a number of DCM steps to achieve extinction, and this is the root cause of the large bias in the DCM method.

\section{Discussion}

We have seen that while the DCM method captures  the correct universal behavior and is exact in the $\Delta t \to 0$ limit, nevertheless it is a very poor tool for quantitative studies of phase diagrams and the like, as it possesses a large systematic bias toward the active state. This is why Martin, et al. saw that the extinction transition point moves to lower values of $\alpha$ with increasing noise, which is physically unacceptable.  

\begin{figure}
\centering
\includegraphics[width=0.8\textwidth]{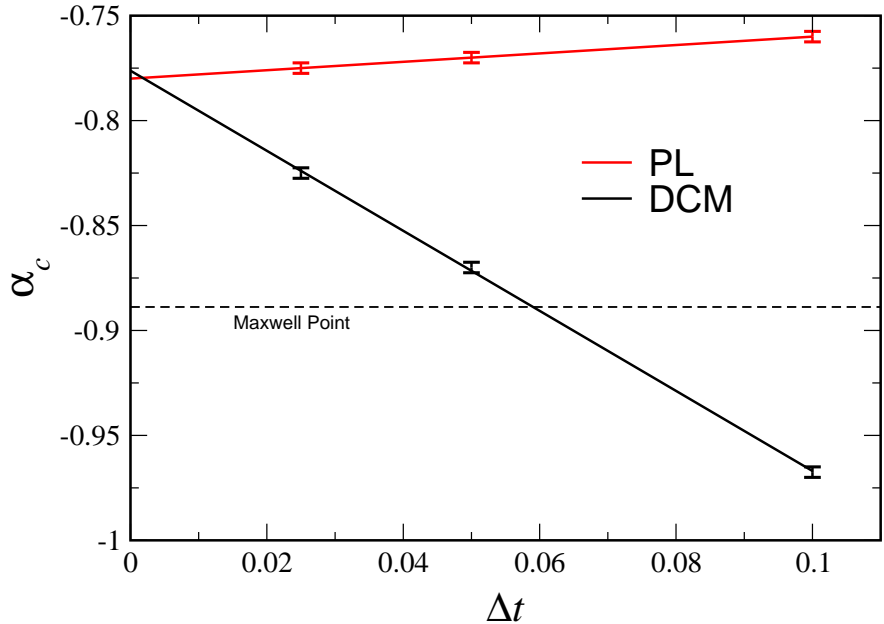}
\caption{The location $\alpha_c$ of the stochastic stall point for the active-inactive phase boundary for the two-dimensional model as a function of $\Delta t$, with $D=1$, $\beta=2$, $\gamma=1$, $\sigma^2=1$.  Also shown is the value of the deterministic Maxwell point, which is crossed by the DCM results as $\Delta t$ is varied. The system size is $2^7\times 2^7$.
The diffusion step of the PL method was done using ADI.}
\label{fig2D}
\end{figure}

To verify this specific point in the 2D scenario considered in Martin, et al., we have measured the location of the stochastic stall point  as a function of $\Delta t$ for the set of parameters singled out by Martin, et al., where the stochastic transition was on the high-stress side of the deterministic Maxwell (stall) point.  As seen in Fig. \ref{fig2D}, and consistent with what we have seen above, the DCM transition point has a strong dependence of $\Delta t$ and indeed crosses over to the correct side of the deterministic Maxwell point in the $\Delta t \to 0$ limit.  The PL transition point has a very much weaker dependence on $\Delta t$ and only slightly overestimates the effect of stochasticity.  We are currently engaged in a detailed study of the phase diagram of the two-dimensional system using the more quantitatively (and qualitatively) reliable PL method. 

Actually, it was quite natural for DCM to incorporate the diffusion into the stochastic term, since in principle the more terms one can handle analytically the better. However, once the external source term is taken as constant during the update step (which is a harmless approximation in the deterministic case), adding the source to the stochastic update actually changes the physics in a major way, but making extinction impossible in a single time state, and this then becomes an unwise move.

Lastly, we point out that  the hybrid method discussed above, implementing the deterministic version of DCM's on-site sink/off-site source breakup of the diffusion operator, which in arbitrary dimension $d$ reads
\begin{equation}
\phi_i''  = \phi_i' e^{-2D\Delta t} + \left(\frac{1}{2d}\sum_{j\in\textit{n.n.}}\phi_j'\right) \left(1 - e^{-2D\Delta t}\right),
\end{equation}
 is computationally the most efficient method, as it is unconditionally stable, and the lower order accuracy (compared to the $d$-dimensional Alternating Direction Implicit~\cite{NumRec} (ADI) method) is swamped by stochastic errors.

\acknowledgments{NMS acknowledges funding from the Israel Science Foundation, grant 1427/15. DAK acknowledges funding from the Israel Science Foundation, grant 1898/17.}

\bibliographystyle{apsrev4-1}
\bibliography{munoz.bib}

\end{document}